08/02/2011

# The extreme physical properties of the CoRoT-7b super-Earth

A. Léger[a,b], O. Grasset[c], B.  Fegley[d], F. Codron[e], A.,  F. Albarede[f], P. Barge[g], R. Barnes[h], P. Cance[c], S.  Carpy[c], F. Catalano[i], C. Cavarroc[a,b], O. Demangeon[a,b], S. Ferraz-Mello[j], P. Gabor[a,b], J.-M. Grießmeier[k], J. Leibacher[a,b,q], G. Libourel[l], A-S. Maurin[l,q], S.N. Raymond[l,q], D. Rouan[n], B. Samuel[a], L. Schaefer[c], J. Schneider[o], P. A. Schuller[a], F. Selsis[l,q], C. Sotin[p].

Affiliations

a. *(Corresponding author)* Institut d'Astrophysique Spatiale, Université Paris-Sud, bât 121, Univ. Paris-Sud, F-91405 Orsay, France;
   e-mail: Alain.Leger@ias.u-psud.fr;  ph: 33 1 69 85 85 80

b. Institut d'Astrophysique Spatiale, CNRS (UMR 8617), bât 121, Univ. Paris-Sud, F-91405 Orsay, France,

c. Université de Nantes, CNRS, Lab de planétologie et Géodynamique, UMR–CNRS 6112, F–44300 Nantes

d. Planetary Chemistry Laboratory McDonnell Center for the Space Sciences, Dep. of Earth and Planetary Sciences, Washington University in St. Louis, USA;

e. Laboratoire de Météorologie Dynamique CNRS/UPMC, T 45-55, E3, 75252 Paris Cedex 05 – France;

f. Ecole Normale Supérieure (LST), 69364 Lyon cedex 7, France; Laboratoire d'Astrophysique de Marseille, Pôle de l'Étoile, 38 r. Frédéric Joliot-Curie, 13388 Marseille cedex 13, France

g. Department of Astronomy, University of Washington, Seattle, WA, 98195-1580, USA;

h. 12 av. Fontaine de la Reine, 92430 Marnes la Coquette, France;

i. IAG-Universidade de Sao Paulo, Brasil;

j. Centre de Biophysique Moléculaire CNRS, r. Charles Sadron, 45071 Orléans cedex 2, France;

k. CRPG-CNRS, 15 Rue Notre-Dame des Pauvres, BP 20, 54501 Vandoeuvre les Nancy, France; e-mail:

l. Université de Bordeaux, Observatoire Aquitain des Sciences de l'Univers, 2 r. de l'Observatoire, BP 89, F-33271 Floirac Cedex, France;

m.  LESIA, UMR 8109 CNRS, Observatoire de Paris, UVSQ, Université Paris-Diderot, 5 pl. J. Janssen, 92195 Meudon, France;

n.  LUTH, UMR 8102 CNRS, Observatoire de Paris-Meudon, 5 pl. J. Janssen, 92195 Meudon, France;

o.  Jet Propulsion Laboratory, California Technology Institute, Pasadena, CA, USA;

p.  National Solar Observatory, 950 N. Cherry Avenue, Tucson, AZ 85719-4933, USA;

q.  CNRS, UMR 5804, Lab. d'Astrophysique de Bordeaux, 2 r. de l'Observatoire, BP 89, F-33271 Floirac Cedex, France

Short title: Physical properties of CoRoT-7b

Abstract:

The search for rocky exoplanets plays an important role in our quest for extra-terrestrial life. Here, we discuss the extreme physical properties possible for the first characterized rocky super-Earth, CoRoT-7b ($R_{pl}$ = 1.58 ± 0.10 $R_{Earth}$, $M_{pl}$ = 6.9 ± 1.2 $M_{Earth}$). It is extremely close to its star ($a$ = 0.0171 AU = 4.48 $R_{st}$), with its spin and orbital rotation likely synchronized. The comparison of its location in the ($M_{pl}$, $R_{pl}$) plane with the predictions of planetary models for different compositions points to an Earth-like composition, even if the error bars of the measured quantities and the partial degeneracy of the models prevent a definitive conclusion. The proximity to its star provides an additional constraint on the model. It implies a high extreme-UV flux and particle wind, and the corresponding efficient erosion of the planetary atmosphere especially for volatile species including water. Consequently, we make the working hypothesis that the planet is rocky with no volatiles in its atmosphere, and derive the physical properties that result. As a consequence, the atmosphere is made of rocky vapours with a very low pressure ($P \leq 1.5$ Pa), no cloud can be sustained, and no thermalisation of the planetary is expected. The dayside is very hot (2474 ± 71 K at the sub-stellar point) while the nightside is very cold (50 to 75 K). The sub-stellar point is as hot as the tungsten filament of an incandescent bulb, resulting in the melting and distillation of silicate rocks and the formation of a lava ocean. These possible features of CoRoT-7b could be common to many small and hot planets, including the recently discovered *Kepler*-10b. They define a new class of objects that we propose to name *"Lava-ocean planets"*.

# 1. Introduction: the CoRoT-7 planetary system

Rocky planets located in the habitable zone (HZ) of their star are our present best candidates for harbouring extra-terrestrial life (see for instance Cockell et al., 2009). As a result, the search for rocky exoplanets is of special importance. Super-Earth candidates, planets with masses 1 - 10 $M_{Earth}$ (Grasset et al., 2009) have recently been discovered (Rivera et al., 2005; Lovis et al., 2009; Bouchy et al., 2009), but only a lower limit for the masses and no radii were measured so that their nature remains ambiguous. The CoRoT mission[1] (Auvergne et al., 2009) and the associated follow-up have detected and characterised the first super-Earth with both radius and mass simultaneously measured, CoRoT-7b (Léger, Rouan, Schneider et al., 2009, thereafter LRS09; Queloz et al., 2009).

Recently, Charbonneau et al. (2009) have discovered a small transiting planet around an M star, GJ1214 b, and measured its radius and mass ($R_{pl}$ = 2.68 $R_{Earth}$, $M_{pl}$ = 6.55 $M_{Earth}$). They interpret their results as being due to a Neptune-like planet with a (H – He) envelope, which would prevent it from being a rocky planet, as opposed to CoRoT-7b as discussed in Sect.2.

The parent star, CoRoT-7 is an active 1-2 Gyr G9V star 150 pc from the Sun, whose observation has recently been carefully reanalysed (Bruntt et al, 2010). The resulting effective temperature, gravity, mass, radius and luminosity are $T_{eff}$ = 5250 ± 60 K, log $g$ = 4.47 ± 0.05, $M_{st}$ = 0.91 ± 0.03 $M_{Sun}$, $R_{st}$ = 0.82 ± 0.04 $R_{Sun}$, and $L$ = 0.48 ± 0.07 $L_{Sun}$, respectively. Its composition is close to that of the Sun, [M/H] = + 0.12 ± 0.06.

The planet, CoRoT-7b, has a radius $R_{pl}$ = 1.58 ± 0.10 $R_{Earth}$, or 10,070 ± 640 km (LRS09 with the stellar parameters of Bruntt et al, 2010). Its orbital period, as measured from the transit light curve, is $P_{orb}$ = 0.85359 ± 3 $10^{-5}$ days. This short orbital period indicates that the planet is very close to its star, at 4.48 times the stellar radius from its centre ($a$ = 0.01707 ± 0.00019 AU = 4.48 ± 0.22 $R_{st}$), with the important implication that spin and orbital rotations are expected to be phase-locked by tidal dissipation (LRS09). The dayside would be continuously irradiated, and the nightside continuously in the dark.

The determination of the planetary mass has led to an intensive radial velocity (RV) campaign, using the HARPS spectrometer (109 measurements and 70 hours of observation), and different analyses of the data. To our knowledge at the date of the 20[th] of January 2011, (1) Queloz et al (2009) made the initial analysis. After revision of CoRoT-7 stellar parameters by Bruntt et al (2010), it yields $M_{pl}$ = 5.2 ± 0.8 $M_{Earth}$; (2) Hatzes et al. (2010) performed another analysis of the data and find a larger mass, $M_{pl}$ = 6.9 ± 1.4 $M_{Earth}$;

---

[1] The CoRoT space mission has been developed and is operated by CNES, with the contribution of Austria, Belgium, Brazil, ESA, Germany, and Spain.

(3) Pont et al. (2010) find a much lower mass with a very large uncertainty, $M_{pl} = 1 - 4\ M_{Earth}$ (1 $\sigma$), but this estimate has recently been questioned by Hatzes et al. (2011); (4) Boisse et al. (2011) find $M_{pl} = 5.7 \pm 2.5\ M_{Earth}$; (5) Ferraz-Mello et al. (2011) find $M_{pl} = 8.0 \pm 1.2\ M_{Earth}$; and (6) Hatzes et al. (2011) find $7.0 \pm 0.5\ M_{Earth}$. In the present state of knowledge, the determination of the mass of CoRoT-7b from the RV measurements of its moderately active host star is delicate. The noise due to the stellar activity dominates the planetary signal and is difficult to remove, a necessary stage to derive the planetary mass. Likely, this situation will recur in the future, when smaller planets will be detected at larger distances from their star by transit observations[2].

However, we have a set of six analyses of the (same) RV data that use different methods to discriminate the planetary signal and the stellar activity. With the exception of the value by Pont et al. (questioned) the five values are reasonably gathered, so we propose to use the median value and the median estimate of the uncertainty:

$$M_{pl} = 6.9 \pm 1.2\ M_{Earth}.$$

In the presence of tidal forces, the shape of the planet is approximately an ellipsoid. We calculate the deviations from a sphere and find: a' - $R_{pl}$ = 66 km, $R_{pl}$ – b' = 19 km, and $R_{pl}$ – c' = 47 km, where a', b' and c' are the ellipsoid semi-axes, and $R_{pl}$ their mean value. These deviations are significantly less than the uncertainty on the mean radius (± 600 km) and can be neglected in the density calculation.

A second planet, CoRoT-7c, was found by Queloz et al. (2009) by analysing the RV data, with a 3.69 day period, a 8.4 $M_{Earth}$ minimum mass, and a 0.045 AU = 11.9 $R_{st}$ semi-major axis. This planet is confirmed by the analysis of Hatzes et al. (2010). Contrary to CoRoT-7b, it does not transit, which is consistent with its larger distance to the star, and consequently its radius remains unknown.

---

[2] With the possible exception of RV measurements in the IR, if the stellar activity has sufficiently less impact in this wavelength domain

## 2. The composition of CoRoT-7b

The simultaneous measurements of the radius and mass of a planet allow us to estimate its composition, at least to some extent. In Fig.1, these quantities for CoRoT-7b are compared to the predictions of models of planetary internal structure for different compositions (Sotin et al., 2007; Fortney et al., 2007; Seager et al., 2007; Valencia et al., 2007a; Grasset et al., 2009). The most probable values of $R_{pl}$ and $M_{pl}$, are located in a domain that crosses the $R_{pl}(M_{pl})$ curve of telluric rocky planets (metallic core + silicates mantle, in terrestrial proportions).

It must be noted that (i) the uncertainties on the measured quantities prevent a definitive conclusion because the domain where the planet lies in the ($R_{pl}$, $M_{pl}$) plane, with a 95% probability (2 $\sigma$), includes planets with a substantial amount of water; (ii) a priori, there is a degeneracy in the $R_{pl}$, $M_{pl}$ diagrams because a planet with a larger dense core and some light materials, e.g. water, in its outer parts can lie at the same position as another planet with a smaller dense core and heavier materials in its outer parts.

Figure 1

There is an additional piece of information to estimate the most probable composition of the planet. CoRoT-7b is located very close to its parent star and therefore subjected to an intense extreme UV flux and particle wind. An efficient erosion of the planetary atmosphere results from thermal and non-thermal escape (Grießmeier 2005; Erkaev et al, 2007; Lamer, 2009; Valencia et al., 2010). The former is especially strong for hydrogen and helium (thermal velocity proportional to (molecular mass)$^{-1/2}$, escape velocity independent of mass), so it is probable that the planet has no (H - He) envelope, unless we observe it at a very special time, just when the remaining mass of this envelope is a minute fraction of the planetary mass.

The case of a significant water content is similar. An upper limit of the escape rate is given by the energy-limited regime where the whole energy deposited by EUV photons and wind particle is converted into escape energy. In this regime, 6 – 7 Earth masses would be eroded at the planet's 1.5 Gyr age (Selsis et al., 2007, Fig.4), so that a small water residual at the present time, $M_{H2O} < 1 M_{earth}$, would also be unlikely. The actual ratio of the energy conversion depends on the cooling processes in the atmosphere, and is unknown. Yelle et al. (2008) and Tian et al. (2008) proposed non-vanishing values, e.g. a few tenths, which would still point to a strong erosion of the atmosphere.

In agreement with Valencia et al. (2010), we conclude that the atmospheric erosion processes for CoRoT-7b are likely to be so efficient that volatile species such as $H_2$, He, $H_2O$ have gone and that only other materials having major reservoirs (> 1 $M_{earth}$) are still present in its atmosphere.

Consequently, we adopt the *working hypothesis that the planet is rocky, with the composition of a dry Earth*, as suggested by both preceding arguments. Several extreme properties of the planet result from this hypothesis and the direct application of the laws of physics and chemistry, as discussed in Sect. 4 to 7.

## 3. Formation and orbital history

### 3.1 Formation

Explaining the presence of planets such as CoRoT-7b is a challenge. Neither Ida and Lin (2008) nor Mordasini et al. (2009a, 2009b)'s models produce such planets without ad hoc tuning of their parameters, e.g. type I migration speed and incorporation of additional phenomena. CoRoT-7b type planets could thus lead to a better understanding of planet formation and migration mechanisms. Possible candidates for CoRoT-7b's formation pathway (Raymond et al., 2008) include: (1) in situ accretion, (2) formation at larger distances followed by inward migration, (3) resonant shepherding during the migration of a giant planet, and (4) photo-evaporation of close-in hot neptunes or hot saturns.

- *Scenario (3)* can probably be ruled out because it would require the existence of a gas giant planet just exterior to a strong mean motion resonance with CoRoT-7b, e.g. 2:1 or 3:2 (Zhou et al, 2005; Fogg & Nelson, 2005; Raymond et al, 2006; Mandell et al., 2007), which is clearly not observed in the HARPS radial velocity data (Queloz et al., 2009).

- *Scenario (1),* given the small amount of mass thought to be available in the inner regions of standard protoplanetary disks, planet formation in this zone requires a large increase in solid density. At least one mechanism exists that could enhance the solid surface density close to the inner disk edge, thought to be located at 0.02-0.05 AU in most cases. Boulder size objects are known to drift very quickly in the nebula due to their strong aerodynamic friction with the gas. This mechanism, responsible for the "meter size barrier" in the growth of the planetesimals, could strongly enrich the inner regions of the disc in solid material, initiating planet growth. The migrating boulders would pile up at the inner edge of the disc where the pressure gradient of the gas reverses, and accumulate in a planet by successive collisions. Therefore, in-situ accretion seems possible.

- *Scenario (2)* is also plausible; CoRoT-7b may have migrated in to its current location via tidal interactions with the protoplanetary disc. In this case, multiple close-in planets should exist with comparable masses in resonant or near-resonant configurations (Terquem and Papaloizou, 2007; Ogihara & Ida, 2009). Given that tidal evolution could have separated a near-resonant configuration (e.g., Barnes et al, 2008; Papaloizou & Terquem, 2009), this is consistent with our current knowledge of the CoRoT-7 system: planets b, c and possibly d.

- *Scenario (4):* could CoRoT-7b have formed as a hot neptune or a hot saturn and been photo-evaporated to its rocky core (Baraffe et al., 2004; Hubbard et al., 2007)? This would imply that the planet's mass was larger in the past and would have raised correspondingly larger stellar tides. Photo-evaporation to the core is effective within $\sim 0.025$-$0.050$ AU for planets less than $\sim 70$ $M_{Earth}$ (Raymond et al., 2008) and could have occurred to CoRoT-7b.

We conclude that scenarios (1), (2) and (4) are all possible in the present state of observational constraints.

## 3.2 History of the planetary orbit and tidal heating

To understand CoRoT-7b's orbital history after the formation process described above ($\sim 1$ Myr), we must consider its *tidal evolution*, the effect of tidal bulges raised on the planet by the star ("planetary tides"), and on the star by the planet ("stellar tides"). For close-in planets, planetary tides reduce the orbital eccentricity, $e$, and semi-major axis, then rapidly lock the spin rotation to the orbital one ($< 1$ Myr). After that tidal locking, there are no longer major planetary tides, and dissipation in the star is the main driver of evolution (Ferraz-Mello et al., 2008).

The basic timescale of the planetary tidal effects is that corresponding to the circularization of the orbit. At the current distance to the star, the circularization time of CoRoT-7b is around 1 Myr. This is the time necessary to reach an eccentricity smaller than a few 0.001. After, the eccentricity continues to decrease, and in 20 Myr it reaches values below $10^{-4}$. These results are obtained assuming for CoRoT-7b a planetary dissipation factor $Q'_{pl}$ (= $3Q/2k$; Carone and Pätzold, 2007) similar to that of the terrestrial planets ($Q'_{pl} \sim 100$, or $Q \sim 20$) that seems adequate for a planet with possibly a liquid ocean in contact with a solid floor (Sect.6).

It is important to note that when spin and orbital rotations are locked, the only remaining tidal heating is due to the (small) eccentricity of the orbit. Its power can be obtained from Fig.2 of Barnes et al (2010). With an eccentricity of a few $10^{-5}$, we obtain a 0.1 W m$^{-2}$ tidal heat flow. This value is inferior to the geothermal heat due to the fossil and radiogenic heating of this large rocky planet (~ 0.4 W m$^{-2}$, Sect.5.4). Then, it seems likely that the tidal heating of the present planet is no more a major process, as opposed to the case of Io.

The past 1.5 Gyr evolution of CoRoT-7b's semi-major axis is calculated in Barnes et al. (2010), assuming that it evolved on a zero-eccentricity orbit under the effect of stellar tides. They find that CoRoT-7b's initial orbital distance should have been between 0.017 and 0.026 AU, 1.5 Gyr ago, the range comes from the large uncertainty in the value of the stellar tidal dissipation function $Q'_{st} = 10^5 - 10^7$.

Looking to the future, CoRoT-7b will continue to spiral inward and heat up. For values $Q'_{st} = 10^5 - 10^6$ the planet is likely to fall into the star and be destroyed in the next 0.2 – 2 Gyr, and for larger values of $Q'_{st}$ the planet should survive longer. Again the large uncertainty on the stellar dissipation factor is a major problem that prevents the derivation of firm conclusions. An improvement of our understanding of the corresponding physical processes is highly desirable.

## 4. Internal structure and thermal history

### 4.1 Internal structure model

A model of the internal structure of CoRoT-7b was first proposed by Wagner et al. (2009) who concluded that it is poorly constrained, due to the large uncertainty on the planetary mass.

Within our hypothesis (Sect.2), and following Grasset et al. (2009), we model its internal structure in the present situation, assuming a given mass and a rocky composition (Si, Mg, Fe, O, Ca, Al, Ni, and S) constrained by the solar-like metallicity of the star. In addition to the overall proportion of elements, the amount of Fe trapped within the silicates relative to that in the metallic core must be fixed, e.g. by fixing the Mg/(Mg + Fe) ratio in silicates. Fortunately, this unknown ratio strongly influences the size of the core, but only weakly the planetary radius, the quantity presently accessible to observation. A variation of the ratio from 0.6 to 1.0 implies a core radius increase from 3500 km to 5480 km, but a planetary radius increase of only 7 %, the growth of the dense metallic core being compensated by the fact that silicates are lighter. For CoRoT-7b, we adopt the same value as for the Earth: 0.9.

The interior temperature profile is of secondary importance for determining the planetary radius, because its effect on densities is much weaker than that of pressure. This profile is difficult to estimate, even on Earth, since it depends on poorly known parameters such as the vigour of convection in the different layers. However, the different profiles used in our modelling (Fig.2) lead to very similar profiles for the density and the planetary radii, 10,094 and 10,100 km for cold and hot thermal profile, respectively. The temperature and density profiles obtained are plotted in Fig.2 and lead to the radius value of $1.66 \pm 0.16$ $R_E$ using $M_{pl} = 6.9 \pm 1.2$ $M_{Earth}$, where the uncertainty is mainly due to that on the measured mass. This computed radius agrees with the measured one ($1.58 \pm 0.10$ $R_{Earth}$), but the present uncertainty on the planetary mass prevents this agreement from being discriminating.

It is worth noting that the mantle must be totally solid, but at its surface (Sect. 5 & 6). As shown by Stixrude et al. (2009) on Earth, the melting temperature of perovskite is probably always more than 1000 K above the geotherm, and a few 100 K above the basalt melting curve. The difference increases with pressure throughout the mantle. This feature must be even stronger on CoRoT-7b since the pressure increases more rapidly with depth than on Earth due to the higher surface gravity ($22.4 \pm 4.1$ vs $9.8$ m s$^{-2}$). Even if the temperature seems very high in the mantle, it is not high enough to allow the silicates to be stable in a liquid state at depth if they are not already molten at the surface (Sect. 6). As a consequence, *there should be no underground oceans, or even large magma reservoirs, on the planetary scale*.

Figure 2

The core represents 11 % of the total volume (Fig.2), a proportion similar to that of the Earth. In our simulations, it is predicted to be composed of liquid metals. Nonetheless, the core may include a solid inner part for the coldest cases, but without significant changes of densities and radius estimates. The large mass of the planet implies a large surface gravity and a large fraction of high-pressure mineral phases of ferromagnesian silicates in the mantle, mainly perovskite, magnesiowüstite, and post-perovskite (Shim, 2008). The olivine–perovskite phase transition occurs at 300 km depth (versus 660 km on Earth), and the upper mantle occupies only 8 % of the total volume (Earth: 30 %). Fig.3 illustrates the resulting structure.

Figure 3

## 4.2 thermal history of the planet

As stated, it is difficult to determine the precise thermal profile inside a planet. Nevertheless, a reasonable overall scenario can be envisaged for the CoRoT-7b history, based on the constraints that are available. Whatever the formation scenario is (Sect. 3.1), and within our hypothesis on its composition, CoRoT-7b presents the characteristics of a solid planet similar to the Earth, and the cooling of its interior during its primordial stage can be understood using common thermal evolution models developed for the Earth. The planet being ~ 5 times more massive than the Earth, its accretion energy is much larger and it is reasonable to suggest that the mantle of CoRoT-7b was very hot (> 3000 K) in its primordial stage, with the additional radioactive heating. The core is considered as a metallic sphere that cools down through convection within the silicate. Tidal heating can be neglected as soon as the planet is phase locked, i.e. for ages larger than 1 Myr (Sect. 3.2).

The following evolution up to 1.5 Gyr is difficult to assess. Thermal evolution depends on the surface temperature (Sect. 5 and 7) because the latter fixes the heat flux that is radiated by the planet. It may also depend slightly on the initial conditions that differ depending on the origin of the planet, but Tozer (1972) showed in a pioneering work that subsolidus convection buffers the slight differences due to primordial states, in a short period of time (few 10 Myr). The fact that the surface temperature differs strongly between the dayside and the nightside (Sect. 5) may be of some importance for the evolution of the planet, but will not change the main facts that are: (i) the solid mantle is globally very hot throughout its whole history, and (ii) its convective motions must be very vigorous because the Rayleigh number of the whole mantle is larger than $10^{11}$, whereas it is ~ $10^8$ for the Earth when the same values of the intensive physical parameters are used[3]. A 3D model of the planet's cooling is required to be more precise.

---

[3] For CoRoT-7b, the estimate of the Rayleigh number, the ratio between natural convection (buoyancy forces) and heat diffusion, $\alpha \, \rho \, g \, \Delta T \, b^3 / (\kappa \, \mu) = 10^{11}$, is obtained for the thermal dilation coefficient $\alpha = 1 \; 10^{-5} \, K^{-1}$, specific mass $\rho = 5000 \, kg/m^3$, gravity $g = 19 \, m \, s^{-2}$, temperature drop from top to bottom of the silicate mantle $\Delta T = 1000 \, K$, the thickness of that mantle $b = 5 \; 10^6 \, m$, the thermal diffusivity $\kappa = 1 \; 10^{-6} \, m/s^2$, and the kinematic viscosity $\mu < 10^{18} \, Pa.s$.

## 5. Surface temperature distribution

On a phase-locked planet, the dayside is continuously irradiated, and the nightside is in the dark. A major surface temperature asymmetry would result unless an efficient heat transport occurs. An upper limit of the possible heat transport by atmospheric winds and ocean currents can be obtained as follows.

### 5.1 Heat transport by atmospheric winds

To estimate whether CoRoT-7b's atmosphere could thermalise the planetary surface, one can compare the maximum heat transfer by the atmosphere and the radiative exchanges between the surface and the exterior.

Under our hypothesis on the absence of significant amounts of volatiles in the atmosphere (Sect.2), the atmospheric pressure is quite low because of the low value of the vapour pressure of refractory materials, molten or solid rocks, which severely limits the heat capacity of the atmosphere.

In Sect.7, thanks to a model of the planet, we make an explicit estimate of this pressure, (maximum of 1.5 Pa at the substellar point). In addition, the speed of low altitude winds is at most the sound velocity because otherwise a large dissipation would occur and slow it down due to the condition $v_{wind} = v_{ground}$ at the surface[4].

Now, these two values imply an upper limit to the possible heat transport. Considering sonic winds that would travel directly from hot to cold regions (no Coriolis effect), and in the favourable case of condensable species, the maximum transport is

$(dE/dt)_{wind} \sim R_{pl} H c P L / (k_B T)$

where $H$ is the atmospheric scale height, $H = k_B T/(\mu g)$, $T$ the local temperature, $\mu$ the mean molecular mass of the gas, $g$ the surface gravity, $c$ the sound velocity, $P$ the surface pressure, $L$ the latent heat of condensation per molecule, and $k_B$ the Boltzmann constant. Typical maximum values for CoRoT-7b are: $R_{pl} = 1.0\ 10^7$ m, $T = 2500$ K, $\mu \sim 35$ amu (Mg, SiO), $g = 24$ m s$^{-2}$, $H = 60$ km, $P = 1.5$ Pa (Sect.7), $c = 1700$ m s$^{-1}$. Using the (large) latent heat of water, $L = 4.1\ 10^4$ J mole$^{-1} = 0.43$ eV mol$^{-1}$, the upper limit for the heat transportation by winds at planetary scale, is $(dE/dt)_{wind} \sim 10^{16}$ W.

The stellar irradiation power at the location of CoRoT-7b is $\sigma\ [(R_{st}/a)^{1/2}\ T_{st}]^4$, or 2.1 MW m$^{-2}$, more than 3 orders of magnitude larger than the solar irradiance at the Earth (1. $10^3$ kW m$^{-2}$),

---

[4] If one considers the acceleration of an atmospheric test volume by pressure, from the highest pressure location (sub-stellar point, $P = 1.5$ Pa) to the nightside ($P = 0$), neglecting friction, an upper maximum speed is reached, $v_{max} = (2\ g\ H)^{1/2} \sim 1000$ m s$^{-1}$. The sound velocity being $c = 1700$ m s$^{-1}$, this upper estimate satisfies the limit $v < c$.

*which stresses the extreme physical environment for this rocky planet.* The power received by the planet, assuming a Bond albedo close to zero (see thereafter), is located on the dayside and is d$E$/d$t)_{rad}$ ~ 1.0 $10^{21}$ W.

The maximum thermal power that the winds could carry is then 5 orders of magnitude lower than the asymmetric irradiation of the planet. In addition, the atmospheric pressure and the corresponding heat transportation are expected to vanish on the nightside of the planet where the temperature drops to very low values (Sect. 5.4), and cannot bring any power there.

We conclude that, under our hypothesis of the absence of volatiles. *Winds can neither (i) significantly change the temperature distribution on the dayside, nor (ii) provide heat to the nightside.*

## 5.2 Heat transport by oceanic streams

An ocean of molten refractory rocks is expected at the surface of the hot regions (Sect.6). The viscosity of liquid rocks being thermally activated, at T ~ 2200 K it is much lower than that of the relatively cool lavas (T ~ 1500 K) from Earth's volcanoes. Its value is 2 – 6 $10^{-2}$ Pa.s (alumina at 2200 K as a proxy; Urbain, 1982), closer to that of liquid water at ~ 20°C ($10^{-3}$ Pa.s; Weast et al., 1989) than to that of Earth's lavas ($10^{+2}$ Pa.s; Mysen and Richet, 2005). Therefore, an active circulation is possible within the ocean, and the question arises of whether it can carry sufficient heat to modify the surface temperature distribution, or not. This would happen if the oceanic streams were able to carry a power comparable to the radiative power at the surface.

An attempt to estimate this heat transport is made in Sect.8, but it is not conclusive; a detailed 3D model of the ocean seems necessary. However, the maximum impact that it could have can be calculated readily. If the circulation within the ocean were extremely efficient for carrying heat, it could make its temperature uniform ($T_{oc}$), and the corresponding extent of the ocean can be calculated.

In the most favourable case where Coriolis forces are negligible, one can consider that the ocean has an axial symmetry around the planet-star direction, and extends from zenith angles 0 (substellar point) to $\theta_m$. The radiative power emitted by the ocean (considered as Lambertian) would be

$$\Phi_{em} = 2\pi\ \varepsilon_2\ \sigma\ T_{oc}^{4}\ (1 - \cos\theta_m)\ R_{pl}^{2}\ ,$$

where $\varepsilon_2$ is the mean emissivity of the liquid lava in the wavelength range of a 2200 K (fusion temperature of the refractory rocks; Sect.6.2) black body emission (maximum at 1.4 $\mu$m), $\sigma$ the Stefan's constant, and $T_{oc}$ the uniform ocean temperature.

The power received by the ocean from the star depends upon its Bond albedo. *Clouds are not expected because the atmosphere is too thin* ($P$ < 1.5 Pa; Sect.7) *to sustain them.* The local planetary albedo is then that of the molten lava. It is $A = 1 - \varepsilon_5$, where $\varepsilon_5$ is the lava absorptivity (= emissivity) now in the wavelength range of the stellar emission, e.g. in a black body approximation at $T_{eff}$ = 5250 K (maximum at 0.6 $\mu$m). The received power reads

$$\Phi_{rec} = \pi\, \varepsilon_5\, \sigma\, T_{st}^{4}\, \sin^2\theta_m\, R_{pl}^{2}\ ,$$

Equating these powers gives the zenith angle of the ocean shore, $\theta_m$, as the solution of the equation

$$f(\theta_m) \equiv \frac{1-\cos(\theta_m)}{\sin^2(\theta_m)} = \frac{R_{st}^2}{2\,a^2}\,\left(\frac{T_{st}}{T_{oc}}\right)^4\,\frac{\varepsilon_5}{\varepsilon_2}$$

The $f(\theta_m)$ function steady increases from 0.5 to 1.0, for $\theta_m$ increasing from 0 to 90°. As expected, the relation yielding $\theta_m$ indicates that larger ocean extensions are obtained for lower ocean temperatures. To determine the maximum extension of the ocean, its temperature and emissivity must be estimated. The chemical composition of the lava ocean is calculated in Sect. 6. We find 87% $Al_2O_3$, 13% CaO, with a liquidus temperature of 2150 K (FACT data base, 2010). The emissivity of pure alumina at its fusion temperature was recently measured as a function of wavelength by Petrov and Vorobyev (2007). They found $\varepsilon_2$(1.4 $\mu$m) = 0.90 ± 0.02, and $\varepsilon_5$(0.6 $\mu$m) = 0.89 ± 0.02. We use this alumina as a proxy of the oceanic material for its emissivity, and find: $\varepsilon_5 / \varepsilon_2$ = 1.00 ± 0.03.

The lowest possible temperature of the ocean is that of the fusion of its material at zero pressure: 2150 K. With the values of $R_{st}$ and $a$, $f(\theta_m)$ is 0.88 and the zenith angle of the ocean shore is $\theta_m$ = 75° ± 2°, corresponding to a 37% coverage of the planet. Because this extension estimate is a maximum, we conclude that *the ocean is limited to the dayside, and its circulation cannot carry heat from the dayside to the nightside.*

In the following, we calculate the temperature distribution on the dayside in a first approximation, where the temperature distribution on the dayside results from the local balance between the absorption of the stellar light and the radiative emission, neglecting a possible heat transport within the ocean. In the future, if a 3D modelling of the oceanic circulation, including the Coriolis effect, were to indicate that this approximation is not valid, a detailed temperature map should be established.

### 5.3 Dayside, surface temperature distribution

If the incoming light is approximated as a parallel beam, in the absence of greenhouse effect (thin atmosphere), the radiative balance gives a temperature (LRS09):

$T_{surf} = (\varepsilon_5/\varepsilon_2)^{1/4} \, (R_{st}/a)^{1/2} \, (\cos\theta)^{1/4} \, T_{st}$ ,

where $a$ is the distance to the star, $\theta$ the azimuth angle with respect to the stellar direction, and $T_{st}$ the effective temperature of the star. As seen previously, for pure $Al_2O_3$, the two emissivities $\varepsilon_5$ and $\varepsilon_2$ are equal ($\pm$ 3%), and we assume that this remains true for the ocean compositions that are dominated by alumina (Sect. 6.1). A surface temperature results that varies from a sub-stellar temperature $T_{sub\text{-}st}$ = 2474 $\pm$ 71 K at $\theta$ = 0, a value similar to that of the filament of an incandescent lamp, down to very low temperatures on the night side (Fig.4, dashed line).

Now, the angular size of the host star as seen from the planet is large (angular diameter, 30°), and must be taken into account when determining precisely the direction of the incoming light, in a second approximation. If $\mathbf{n}$ is the unit vector normal to the surface at a given point of the planet, the irradiance coming from an element $\mathbf{d}\omega$ of the stellar surface is proportional to $d\Omega = \mathbf{n}\cdot\mathbf{d}\omega$ . Using relations resulting from the geometry of the system, the integral over the surface of the stellar irradiance is

$$\Omega(\theta) = \int_{-R_2}^{min(h,R_2)} \cos(\arctan[D\sin(\theta)/(h-z)]) \times 2(R_2^2 - z^2)^{1/2}/[(D\sin(\theta))^2 + (h-z)^2] \ dz$$

If the star were distant ($D \gg R_{st}$), one would find the classical result: $\Omega_0 = \pi R_{st}^2/D^2$, at the sub-stellar point. The temperature resulting from this irradiation is plotted in Fig.4 (solid line).

The difference between the two temperature distributions is small in highly irradiated regions (e.g. unchanged $T_{sub\text{-}st}$), but important in the terminator region (penumbra effect). In particular the nightside begins at a zenith angle of $\theta$ = 102.7° $\pm$ 0.7°, instead of $\theta$ = 90° in the parallel beam approximation.

*5.4 Nightside heating and resulting temperature*

The temperature of the nightside results from the balance between radiative cooling and different possible heating mechanisms. The latter are:

- the geothermal flux and possibly the tidal flux. Using our model of CoRoT-7b's cooling, we can estimate the radioactive heating and the primordial accretion energy output. The ratio of the corresponding output fluxes, the Urey number, is found to vary from 20% for the early planet, to 90% for the 1.5 Gyr planet. The present geothermal flux is ~ 0.4 W m$^{-2}$, several times the value for the present Earth (~ 0.075 W m$^{-2}$). The tidal heating is probably less (~ 0.1 mW m$^{-2}$; Sect.3.2) as a result of the rotation phase locking;

- a heating by the stellar corona plasma that irradiates the nightside. The stellar corona can be described as a thin atmosphere at ~ 10$^6$ K and a radiative emission of $\Lambda_0 = 10^2$ W m$^{-2}$ (Withbroe et al., 1977). At the planet's distance from the star ($d = 2 \times 10^6$ km) the plasma column density is reduced by a factor e$^{-d/h}$, where $h$ is the coronal scale height. Using $\mu = 1$ amu, $g = 340$ m s$^{-2}$, $h \sim 5 \times 10^4$ km, the resulting heating is $\Lambda_{plasma} \sim \Lambda_0$ e$^{-d/h} \sim 4 \times 10^{-6}$ W m$^{-2}$, which is negligible.

As we have shown that liquid lava is not expected to flow over the nightside (Sect.5.2), neither is an underground ocean that would have increased the geothermal flux, the total heating on that side is probably dominated by the usual geothermal heating and can be estimated to 0.5 ± 0.2 W m$^{-2}$ for CoRoT-7b. The balance between this flux and the radiative emission of the surface, assuming a unit emissivity, would lead to a surface temperature of $T_{surf} \sim 55 \pm 5$ K. Now, the albedo of the nightside could be high as the result of deposition of frost material provided by icy planetesimals (comets). If a mean albedo of $A = 0.8$ (fresh snow) is considered as a maximum, *the nightside surface temperature is in the range 50-75 K.*

# 6. A permanent magma ocean

*6.1 Building up of an lava ocean*

Within our hypothesis, the planetary atmosphere is the result of the vaporization of the surface material, depending on the local temperature. Using the MAGMA code (Fegley and Cameron, 1987; Schaefer and Fegley, 2004) we can compute the evolution of the oceanic composition with time, starting from a silicate composition (Schaefer and Fegley, 2009). Fig.5 shows its radical changes as evaporation proceeds, the abundance of refractory species ($MgO$, $SiO_2$, then $TiO_2$, $CaO$, $Al_2O_3$) increases with time, and the total vapour pressure decreases (Fig.6). Eventually, at high vaporized fractions ($F_{vap} > 0.95$), the composition of the refractory residue remains stable, with mole fractions of 13% for $CaO$ and 87% for $Al_2O_3$ at 2200 K (close to the composition $CaO.6Al_2O_3$), which we propose for the composition of the ocean after a 1.5 Gyr evolution.

Figure 5

Figure 6

The evaporative mass transport, if irreversible, can change the composition of the surface layers of the planet. Assuming an evaporation efficiency coefficient of 0.1 (Schaefer and Fegley, 2004), the time required for evaporating a 100 km layer is $\tau$ / 3 $10^5$ yr = ($P$ / 1 Pa)$^{-1}$. For pressures of about one Pascal, it is much shorter than the planetary age, indicating that thousands of kilometres can be evaporated during the planet's lifetime.

When the vapour condensation occurs onto the ocean, flows can bring back the volatiles to the depleted area, but this is no longer true when condensation occurs onto the continent, and irreversible transport of material from the ocean to the continent occurs. The resulting reduction of the oceanic volume leads to the dissolution of the oceanic floor and the evolution of its composition that we have estimated. The loaded continent progressively sinks into the mantle as its roots probably dissolve. In the solid mantle, a silicate transport, at low velocities, from the continent roots to the ocean floor may close the circulation of materials. Plate tectonics (Valencia et al. 2007b) may also play an important role in the transport of materials.

## 6.2 Determination of the location of the ocean-shore

The ocean shore is the location on the planetary surface where the solidification of the molten rocks occurs. If crystallization happens close to thermodynamic equilibrium conditions, crystals will grow to rather large sizes, and the density difference between solid and liquid phases will lead to an efficient differentiation: the solids, denser than the liquid, sink down. When the surface temperature reaches that of the local *magma liquidus,* condensation occurs and pieces of rock sink, and the ocean floor rises up to the surface, building the ocean shore. Therefore, we consider that the ocean shore *is located where the surface temperature reaches the liquidus temperature of the local magma*. For the composition that we propose, $(CaO)_{13\%}.(Al_2O_3)_{87\%}$, this happens at $T$ = 2200 ± 20 K (FACT data base, 2010), and zenith angle $\theta$ = 51° ± 5.3°. For an oceanic composition with half as much CaO, the temperature would be 2270 K, and the zenith angle $\theta \sim 45°$.

## *6.3 Determination of the ocean-depth*

The bottom of the ocean is the place where the melting temperature, $T_{melt}(P)$, meets the local temperature. A linear estimate of the melting curve, $T_{melt}(P)$, can be obtained using the Clapeyron equation, $(dP/dT)_{melt} = \Delta H / (T \Delta V)$, where $\Delta H$ is the enthalpy of melting, and $\Delta V$ the associated change in volume. Using values for alumina, the main component of the ocean, one reads $\Delta H$ = 111 ± 4 kJ mol$^{-1}$ (Chase et al., 1985) and $\Delta V_{Al2O3}$ = 0.70 g cm$^{-3}$ the difference between the densities of corundum and molten alumina (Fiquet et al., 1999). The resulting dependence on pressure, $P = \rho\, g\, h$, of $T_{melt}$ is

$T_{melt}$(K) = 2200 + 1.25 10$^{-7}P$(Pa)

If the temperature profile is an adiabat, it reads: $T_{ad}$ (K) = $T_{surf}$ + *1.3 10$^{-8}$* $P$(Pa). With $T_{surf}$ = 2500 K, $\rho$ = 3.05 g cm$^{-3}$, and $g$ = 22 m s$^{-2}$, *the ocean-depth would be 45 km, at the substellar-point.* If the temperature at the ocean floor were lower, e.g. as the result of lava circulation, the ocean would be shallower. Only a 3D model can give a final description.

## 7. Planetary atmosphere

The atmosphere is the result of the vaporisation/sublimation of rocks at the planetary surface. The local equilibrium vapour pressure results from the set of curves of Fig.6 for high vaporization fractions ($F_{vap}$ = 0.95 - 1.00) at the temperature determined in Sect.5.3. The total pressure varies from 1.5 Pa at the substellar point ($T$ = 2500 K), to 3 10$^{-2}$ Pa at the ocean shore ($T$ = 2200 K), and zero (<< 10$^{-10}$ Pa) on the nightside ($T \sim$ 50 K).

An important question is how far is the actual atmospheric pressure from that given by this local equilibrium, $P = P_{sat}(T)$. An upper limit of that difference can be estimated as follows. The temperature gradient leads to a pressure gradient and generates wind. These winds produce local thermodynamic imbalances due to the exchange of gaseous material with the neighbouring regions. An increase of the local pressure drives an increase of the impinging flux onto the surface, which tends to restore the equilibrium. Considering an annulus at zenith angle θ, with a width Δθ, *an upper limit* of the input flux due to winds is obtained assuming that incoming winds at the hot border blow at the sound velocity, and that outgoing winds at the cold border are negligible. The corresponding incoming flux is $(\Phi_{wind})_{max} = 2\pi R_{pl} (\sin\theta) H c n$ , where $n$ is the atmospheric particle density. One can compute the width $\Delta\theta$ needed for that flux to be compensated by the increase of the impinging flux due to excess pressure, $\Delta P$. Requiring that the deviation $\Delta P/P_{vap}$ is 10%, we find that the maximum zenith widths $\Delta\theta$ are 0.90°, 0.85° and 0.80°, for mean zenith angles $\theta$ of 10°, 30°, and 50°, respectively. Considering that these values of $\Delta\theta$ are small compared to the corresponding $\theta$ values, we conclude that *the local atmospheric pressure is the vapour pressure of rocks at the local temperature*, but for a lag inferior to one degree in zenith angle, which we neglect. The resulting pressure distribution is presented in Fig.7.

Figure 7

## 8. Possible oceanic circulation

The low pressure of the atmosphere of CoRoT-7b implies that wind-driven circulation in the lava ocean is negligible. The strong contrasts in radiative equilibrium temperature could however lead to a circulation forced by horizontal density gradients. To estimate the strength of these currents, we consider a vein of thickness $\mathcal{H}$ under the surface, whose temperature $T(x)$ varies in one horizontal direction. The surface of the ocean is at a height $h$ above its mean level $z = 0$. The hydrostatic pressure within the ocean is given by:

$$P(x, z) = \rho g(z + h)$$

The horizontal equation of motion for a stationary flow is:

$$\varepsilon \vec{V} + (\vec{V} \cdot \vec{\nabla})\vec{V} + f\vec{k} \times \vec{V} = -\frac{1}{\rho}\vec{\nabla}P$$

where the $\vec{V}$ term is the horizontal velocity relative to the planet's surface, $\varepsilon$ the inverse timescale of dissipative processes and $f = 2\omega \sin(latitude)$ is the Coriolis factor, with $\vec{k}$ the unit vector in the vertical direction.

The large spatial scale of the ocean implies that the non-linear advective term, $\vec{V} \cdot \vec{\nabla}$, has a very long timescale (of the order of $V/L$, with $L$ the horizontal extent of the ocean, $V/L \sim 10^{-7}$ s$^{-1}$ for $V = 1$m s$^{-1}$), compared to the dissipative and the Coriolis factors whose typical values are $\varepsilon = 10^{-5}$ s$^{-1}$ (see below) and $f = 10^{-4}$ s$^{-1}$ (for a rotational period of 0.85 day, away from the equator). Therefore, this non-linear term can be neglected (Pedlosky, 1996).

Setting the Coriolis force aside for now, the horizontal velocity $u$ in the $x$ direction results from a balance between the pressure-gradient force and the dissipation:

$$\varepsilon u = -\frac{1}{\rho}\partial_x P = -g\partial_x h - gz\frac{\partial_x \rho}{\rho}$$

We assumed that relative density changes are small, and $h \ll \mathcal{H}$. Taking the velocity $u$ to cancel at the bottom of the vein, $z = \mathcal{H}$, we get a velocity, maximum at the surface of the ocean, equal to

$$u = -\frac{1}{\varepsilon}g(\mathcal{H} - z)\beta\partial_x T$$

where $\beta$ is the dilatation coefficient. The mass flux integrated in a vein with thickness $\mathcal{H}$ is

$$F_M = -\frac{1}{2\varepsilon}g\mathcal{H}^2\beta\partial_x T$$

At the point where the mass flux is maximal, a heat flux divergence corresponding to a heat delivery,

$$Q = \rho C_P F_M \partial_x T$$

can be computed and compared to the radiative fluxes at the surface. Using somewhat arbitrarily a dissipative timescale of about one day ($\varepsilon = 10^{-5}$ s$^{-1}$) similar to that in terrestrial oceans (Pickard & Emery, 1990; Pedlosky, 1996), and the radiative equilibrium temperature gradient, a vein depth $\mathcal{H}$ of 100 m [10 km] yields a heating of $Q = 2.1\ 10^4$ W m$^{-2}$ [2.1 $10^6$ W m$^{-2}$], with a maximum current speed of 2.2 m s$^{-1}$. For comparison, a change in surface temperature from 2250 K to 2350 K brings a difference of emission of 2.8 $10^5$ W m$^{-2}$. It thus seems that the capability of ocean currents to modify the surface temperature distribution and the ocean extension significantly, fully relies on the thickness $\mathcal{H}$ of current vein.

A strong circulation in the surface layer necessitates a compensating return flow at depth, which would occur at the colder temperature of the sinking lava, close to the minimum temperature at the surface. The colder lava would then ascend back under the hottest surface, tending to reduce the depth $\mathcal{H}$. Maintaining a deep hot surface layer in the presence of a strong convective cell thus requires a very strong vertical mixing, as the ocean is heated from above. It is not clear what mechanism could make this happen.

As mentioned, the convection and its associated reduction of the temperature at the bottom of the ocean for low azimuth angles would significantly reduce the depth of the ocean. A depth temperature of 2300 K would make the lava solidify at about 14 km.

The Coriolis force was not taken into account in the above calculations. Given the high rotation rate of CoRoT-7b it is a dominant term in the equation of motion far from the equatorial region, with the Coriolis factor $f$ of the order of $10^{-4}$ s$^{-1}$ in the mid latitudes. The main balance will then be between the pressure gradient and Coriolis forces, yielding currents about an order of magnitude weaker and oriented parallel to the isobars. The heat transport by ocean currents will thus be strongly inhibited by the presence of the Coriolis force, and the region showing possible changes in the temperature distribution is restricted to the equator (crossing the sub-stellar point) where the Coriolis factor vanishes. This justifies the two-dimensional aspect of the present calculations, as the ocean circulation will be mostly in the east-west direction.

A more precise determination of the shape and temperature distribution of the lava ocean, as well as its depth, requires a 3D model of the circulation with several hypotheses on the strength of the vertical mixing, which is beyond the scope of the present paper. In any case, as shown in Sect.5.2, this possible oceanic circulation cannot bring heat to the nightside and change its surface temperature.

# 9. Conclusion

We have attempted to describe the expected physical properties of CoRoT-7b, the first super-Earth discovered with radius and mass measured simultaneously, and the key feature of being very close to its star (0.017 AU, or 4.5 stellar radii). The radius is measured with a fair accuracy (± 6%) by transits, and its mass with a ± 14% accuracy by RV data analyses. Considering the additional constraint of the expected rapid erosion of any volatile in the atmosphere, we made the working hypothesis for this composition of a dry earth without atmospheric volatiles. This hypothesis is made plausible by the fit between observations and modelling of planetary interiors, although it does derive unambiguously from them.

In the future, we may expect an improvement in the mass estimate that could qualify or falsify this hypothesis, but this is not certain because it does not rely on the accuracy of the RV measurements but on their analysis that is made difficult by the noise introduced by the stellar activity. This situation will likely repeat in the future when super-Earths further away from their stars, e.g. in their habitable zones, will be detected by transits, because the corresponding planetary part of the RV signal will decrease as ($\propto a^{-1/2}$), typically by a factor 8 for $a$ changing from 0.017 AU to 1 AU, even more for planets further away in the habitable zone.

Within the framework of that hypothesis, we derive the expected physical properties of the planet. We find that its extreme proximity to its star implies very special features. The most striking one results from the probable phase locking of its spin-rotation with its orbital-rotation by tidal dissipation, which implies *a major dichotomy between a hot dayside ($T_{max} \sim 2500$ K) with an ocean of molten rocks, and a cold nightside* ($T$ = 50-75 K). The atmosphere is made of rock vapours with a pressure that we estimate to vary from low values (~ 1 Pa) at the sub-stellar point, to zero on the night side. This low pressure prevents any cloud to be sustained. Possible formation and evolution scenarios for its orbit and internal structure have also be studied. We cannot claim that these are ''the'' properties of CoRoT-7b, but we have built a self-consistent model for it that is compatible with the parameters derived from all available pieces of observations.

Figure 8 shows an artist view of the planet that illustrates several of the features derived in the paper. In particular, we think that the large radiative energy exchanges lead to an intense turnover of materials and to regular frontiers between the lava-ocean and the continent, as opposed to indented water-ocean shores on temperate planets.

These possible features of CoRoT-7b should be common to many small, hot planets, including at present HD 10180b[5], KOI-377.03[6], and *Kepler-10b*[7]. This last planet is curiously analogous to CoRoT-7b and *our physical model fully applies to it* (Table 1).

They define a new class of objects that we propose to name *"Lava-ocean planets"*.

Table 1

**Acknowledgements**

The authors are grateful to Pascal Bordé (IAS) for fruitful discussions on the statistical analysis of the data. We also thank the Programme National de Planétologie of CNRS for its financial support in 2009 and 2010.

---

[5] According to Lovis et al. (2010), the parameters of HD 10180b are $m \sin(i)$ = 1.35 $M_{Earth}$, $P$ = 1.18 day, $a$ = 0.022 AU. The star being a G1V, if the planet is rocky with no thick atmosphere, and phase locked, its substellar temperature would be 2650 K and an ocean of molten rocks would be present at its surface.

[6] According to Holman et al (2010), this object could be a super-Earth, 1.5 $R_{Earth}$ in radius and orbiting around a solar type star in 1.59 days. Under the same hypotheses as for HD 10180b, its substellar temperature would be 2400 K and it would also harbour an ocean of molten rock.

[7] According to Batalha et al (2011), this object is a rocky super-Earth, at the same absolute distance as CoRoT-7b from a somewhat hotter star (5630 K). Under the same hypotheses as for CoRoT-7b, its sub-stellar temperature is 3040 K and it would harbour an ocean of molten rock (see Table 1).

# Table 1

| Parameter | Corot-7b | Note | Kepler-10b | Note |
|---|---|---|---|---|
| $T_{st}$ (K) | $5250 \pm 60$ | (a) | $5627 \pm 44$ | (e) |
| $R_{st}$ ($R_{Sun}$) | $0.82 \pm 0.04$ | (a) | $1.056 \pm 0.021$ | (e) |
| $L_{st}$ ($L_{Sun}$) | $0.48 \pm 0.07$ | (a) | $1.004 \pm 0.059$ | (e) |
| $M_{st}$ ($M_{Sun}$) | $0.91 \pm 0.03$ | (a) | $0.895 \pm 0.060$ | (e) |
| $age$ (Gyr) | $1.2 - 2.3$ | (b) | $11.9 \pm 4.5$ | (e) |
| $mag$ | V=11.7, R=11.3 | (a) | Kepl=10.96 | (e) |
| $Sp\ Type$ | G9V | (a) | ~ G3V | |
| $\Delta F/F$ (ppm) | $335 \pm 12$ | (b) | $152 \pm 4$ | (e) |
| $tr.\ dur.$ (h) | $1.25 \pm 0.05$ | (b) | $1.81 \pm 0.02$ | (e) |
| $b$ | $0.61 \pm 0.06$ | (b) | $0.30 \pm 0.08$ | (e) |
| $i$ (°) | $80.1 \pm 0.3$ | | | |
| $P_{orb}$ (day) | $0.85359 \pm 3\ 10^{-5}$ | (b) | $0.837495 \pm 5\ 10^{-6}$ | |
| $R_{pl}$ ($R_{Earth}$) | $1.58 \pm 0.10$ | (b) | $1.416 \pm 0.034$ | (e) |
| $M_{pl}$ ($M_{Earth}$) | $6.9 \pm 1.2$ | (c) | $4.56 \pm 1.23$ | (e) |
| $\rho$ (g cm$^{-3}$) | $9.7 \pm 1.9$ | (a) + (c) | $8.8 \pm 2.5$ | (e) |
| $a$ (AU) | $0.01707 \pm 0.00019$ | (b) | $0.01684 \pm 0.00033$ | (e) |
| $a/R_{st}$ | $4.48 \pm 0.22$ | (a) + (b) | $3.43 \pm 0.10$ | (e) |
| $F_{st}$ (MWm$^{-2}$) | $2.14 \pm 0.24$ | (d) | $4.87 \pm 0.33$ | (e) + (d) |
| $T_{sub\text{-}st}$ (K) | $2474 \pm 71$ | (d) | $3038 \pm 51$ | (e) + (d) |
| $\theta_{lava}$ (°) | $51.3 \pm 5.3$ | (d) | $73.9 \pm 1.1$ | (e) + (d) |
| $\theta_{day\text{-}night}$ (°) | $102.7 \pm 0.7$ | (d) | $106.7 \pm 0.5$ | (e) + (d) |
| $2\ \theta_{st/pl}$ (°) | 25.8 | (d) | 33.9 | (e) |

*Table 1: Parameters for the CoRoT-7b and Kepler-10b planets and their parent stars, as deduced from observations and the present modelling. $F_{st}$ is the stellar flux at the planet location and, as deduced from our model, $T_{sub\text{-}st}$ is the highest temperature at the surface of the planet (sub-stellar point), $\theta_{lava}$ the angular spreading of the lava ocean, $\theta_{day\text{-}night}$ the position of the day-night frontier (dark limit of penumbra), and $\theta_{st/pl}$ the angle under which the star is seen from the planet. Notes: (a) Bruntt et al. (2010);*

*(b) Léger, Rouan, Schneider et al. (2009) renormalized by stellar parameters from Bruntt et al. (2010); (c) median of the five different mass determinations of CoRoT-7b (see text); (d) present modelling of the physical features of both planets; (e) Batalha et al. (2011).*

# Figure 1

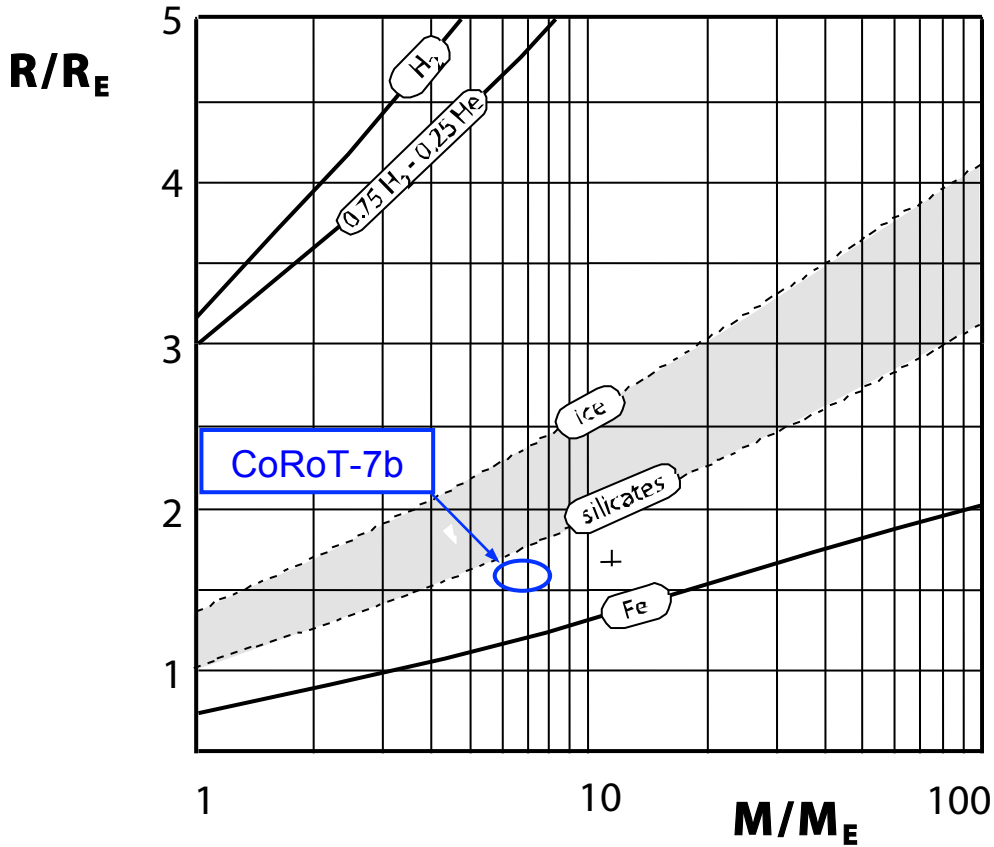

Figure 1: *Radius of planets with different compositions calculated as a function of their mass, adapted from Grasset et al. (2009). Curves labelled [Fe], [silicates], [ices], [H2-He] correspond to planets made of pure Fe, silicates and metallic core, water ice, and pure H2-He gas, respectively. The grey area corresponds to planets with both silicates and water. The domain of the estimated $M_{pl}$ and $R_{pl}$ parameters of CoRoT-7b is represented as an ellipse (68% probability). The radius determination results from the transit measurements (LRS09) and the analysis of the stellar parameters by Bruntt et al. (2010). The mass is a median value of five different estimates from the RV data (see text). The hypothesis of a dry Earth-like composition made in the present paper is pointed out by the position of CoRoT-7b in the R(M) plane but cannot be considered as definitely established, in particular considering possible degeneracies of models e.g. a larger metallic core and a significant amount of water that can mimic the proposed composition. For planets very close to their star, e.g. CoRoT-7b and Kepler-10b (Batalha et al., 2011), the expected harsh EUV radiation and particle wind should produce a fast erosion of any water atmosphere, and this provides an additional piece of information in favour of that hypothesis. The presently determined location even points to a somewhat larger metallic core, but error bars prevent solid conclusions.*

# Figure 2

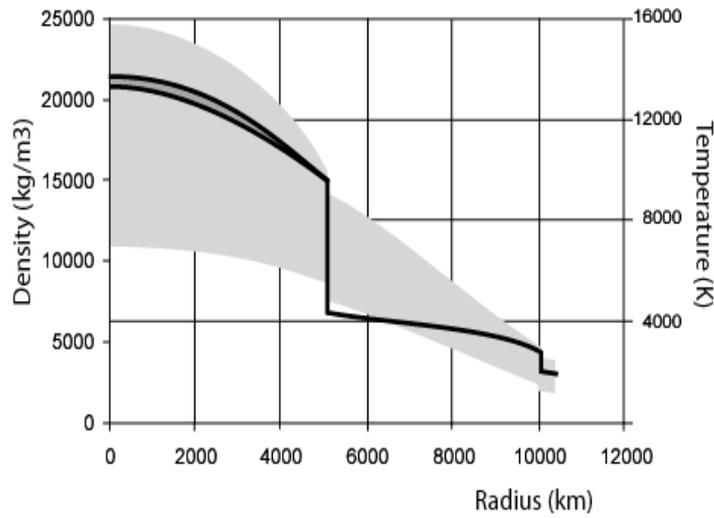

*Figure 2* : *Possible temperature (grey area) and density (dark curves) profiles used for computing the internal structure of CoRoT-7b. Several temperature profiles have been tested within the grey area without significant effects on the position of the layers and the size of the planet ($\Delta R_{pl}/R_{pl} < 10^{-3}$). At this scale, density variations due to temperature contrasts are visible in the core (curve splits) but not in the mantle. In each layer, temperature profiles are assumed adiabatic, and fixed temperature jumps, similar to those encountered on Earth, are imposed at the boundaries. Within each layer, density variations are mostly due to the pressure increase (~ 630 GPa at the core-mantle interface, and ~ 1.7 TPa at the centre).*

# Figure 3

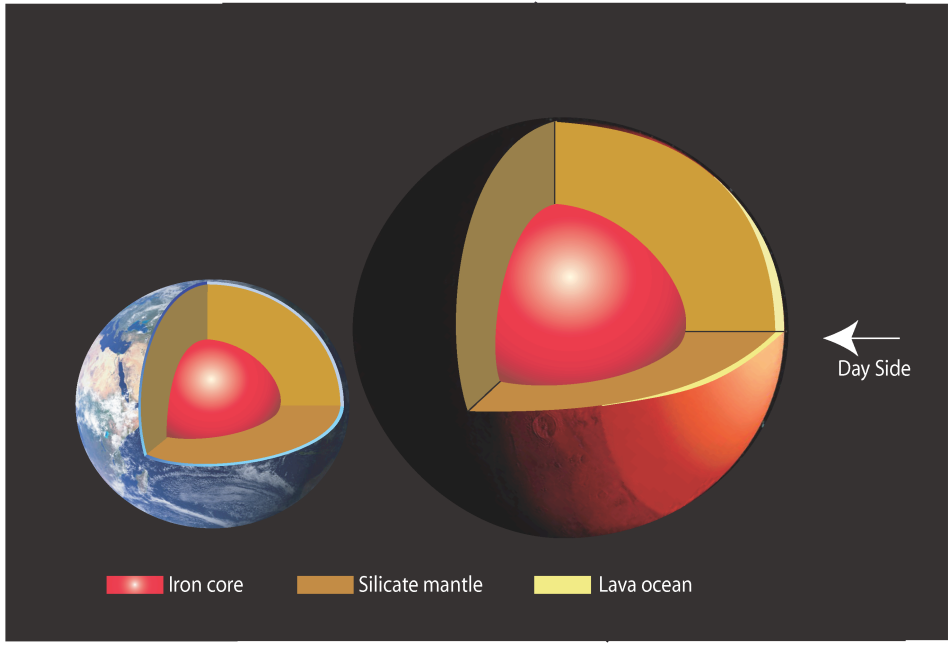

Iron core    Silicate mantle    Lava ocean

**_Figure 3_**: *Model internal structure of CoRoT-7b compared to Earth. Scaling of the different layers has been respected except for the magma ocean thickness, which has been increased in order to be visible.*

# Figure 4

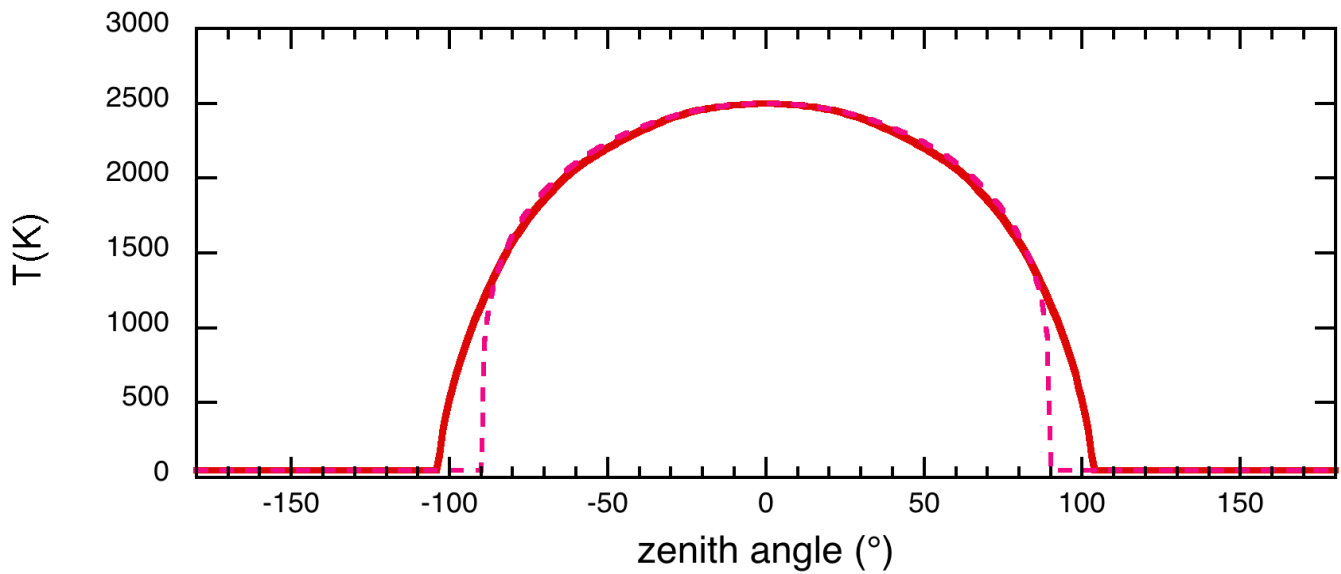

<u>**Figure 4**</u>: *Surface temperature, for a model of CoRoT-7b with a thin atmosphere: (i) within the approximation of a parallel incident beam (dashed line); (ii) taking into account the actual angular size of the star and the resulting penumbra effect (full line).*

# Figure 5

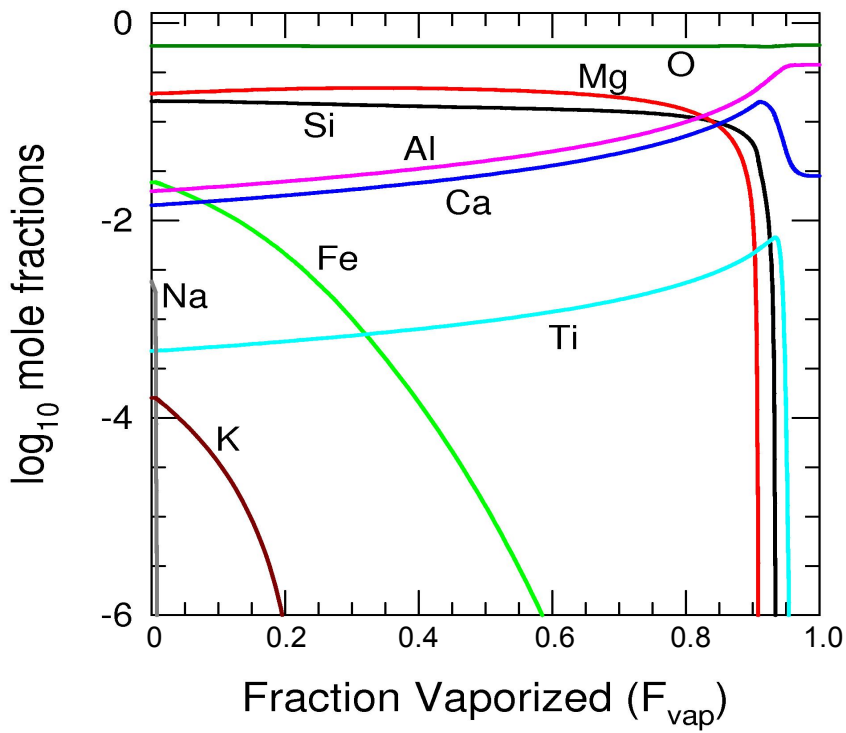

**Figure 5**: *atomic composition of a finite silicate sample, evaporating at constant temperature (T = 2200 K), versus its evaporated fraction, $F_{vap}$, using the thermodynamic code MAGMA (Fegley and Cameron, 1987; Shaefer and Fegley, 2004). As time progresses, the volatile species evaporate, enriching the condensed phase in refractory species. Eventually, at high vaporized fractions, $F_{vap} = 0.96 – 1.00$, the molar composition of the residue is almost constant, [CaO] = 0.13, $[Al_2O_3]$ = 0.87, which is proposed as the composition of the ocean after 1.5 Gyr.*

**Figure 6**

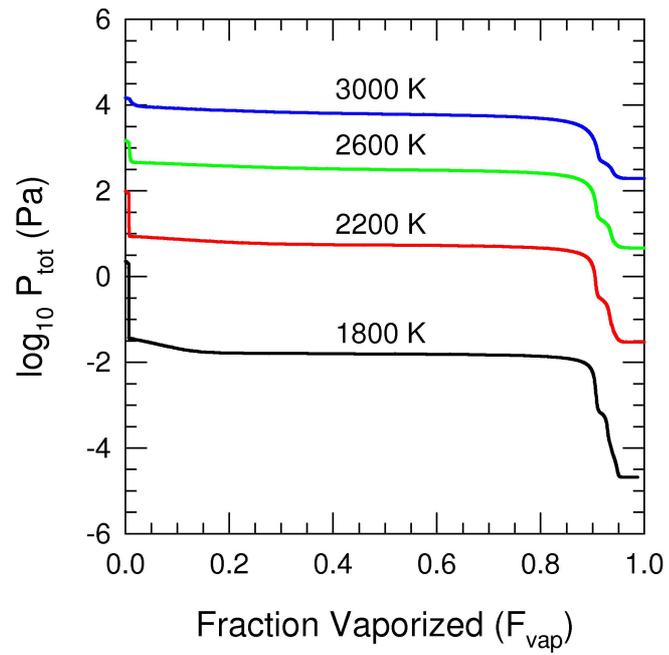

_Figure 6_: total vapour pressure above silicate samples analogous to that used in Fig.5, versus their evaporated fraction, at different temperatures. As expected, with increasing evaporated fraction, the vapour pressures decrease, and the residues become more refractory.

# Figure 7

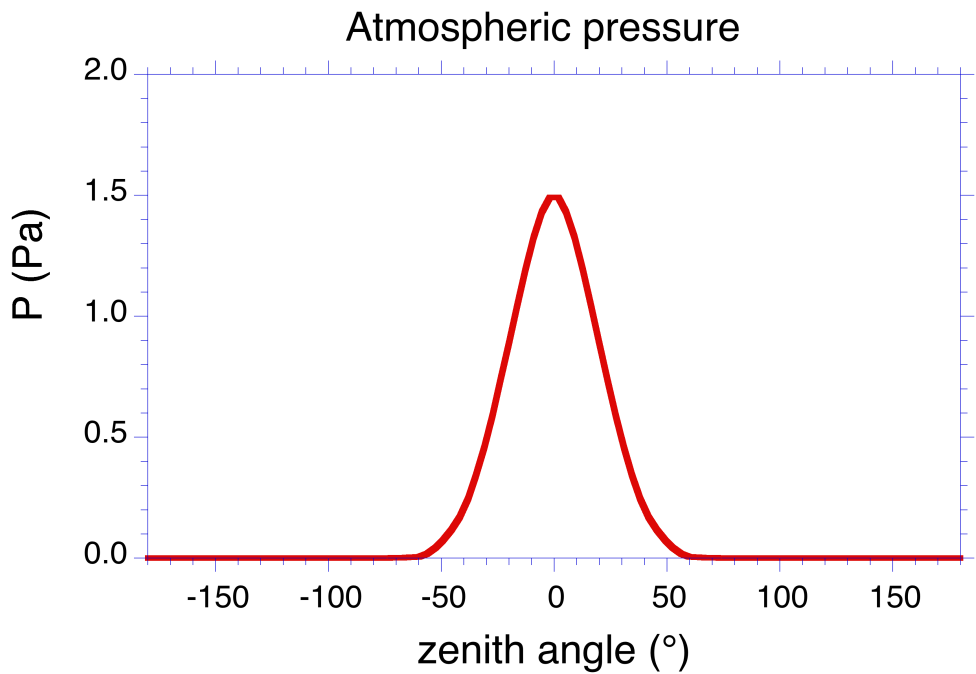

*Figure 7*: Atmospheric pressure resulting from the vaporisation/sublimation of rocks. To a good approximation, it is the vapour pressure of the rocks (Fig.6) at the local temperature (Fig.4).

# Figure 8

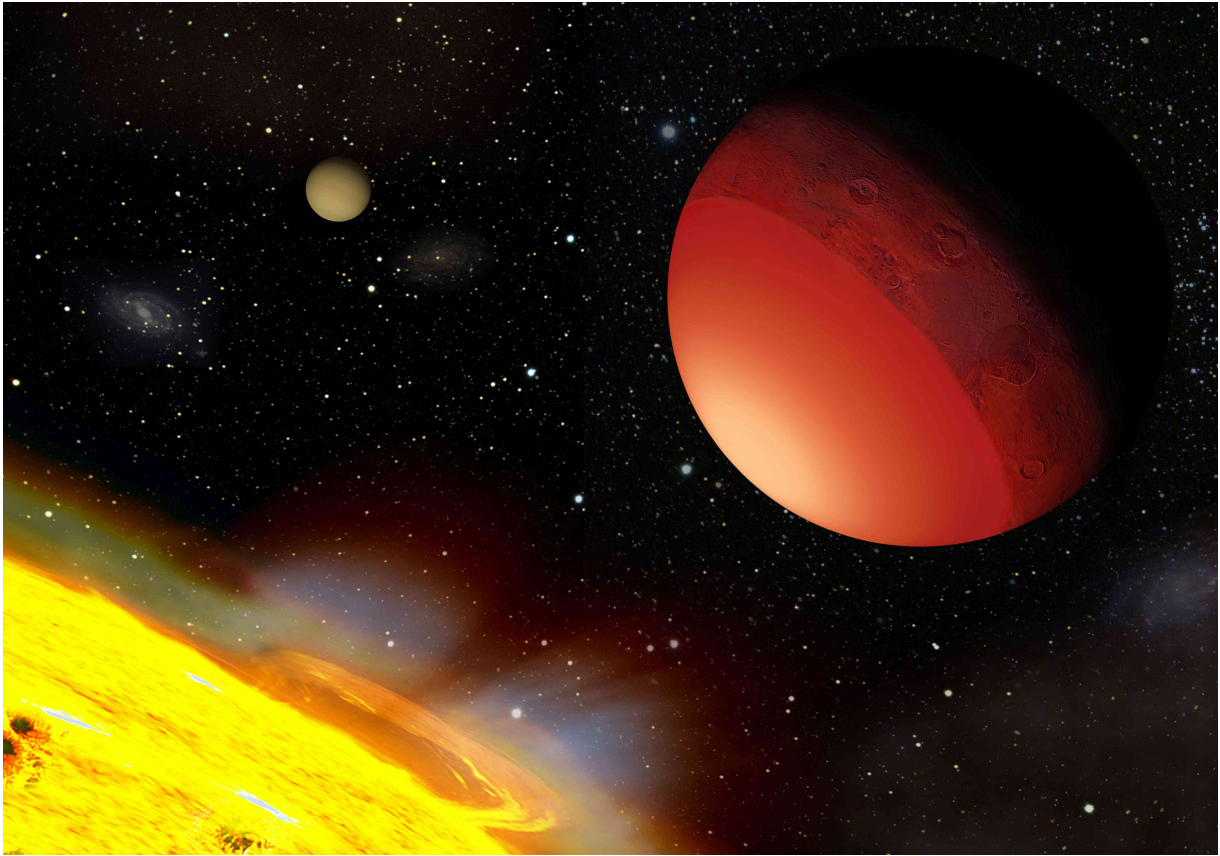

<u>Figure 8</u>: Artist view of the CoRoT-7 system. Stellar wind and coronal mass ejections are schematized at the surface of the active star. A detailed view of CoRoT-7b, is represented, illustrating the main planetary features derived in the paper, including the temperature distribution, the presence of a featureless ocean of molten rocks and a structured continent of solid rocks. In our current understanding of planetary system formation and evolution, the presence of meteorites seems likely, and we represent impact craters on this continent. For the sake of visibility, the planetary size and planet-star distance are not to scale, nor is the relative luminosity of the planet and the star. The second planet of the system, CoRoT-7c, is visible in the distance.